\documentclass[twocolumn]{aastex631} 
\usepackage{graphicx}

\usepackage{xcolor}

\newcommand{\1}{WNTR24-cua}
\newcommand{\2}{WNTR24-egv}

\begin{document}

\title{Multi-color characterization of optically invisible FU Orionis-type outbursts: \\ Demonstration and prospects for the WINTER survey}

\correspondingauthor{Danielle Frostig}
\email{danielle.frostig@cfa.harvard.edu}

\author[0000-0002-7197-9004]{Danielle Frostig}
\affil{Center for Astrophysics $\vert$ Harvard \& Smithsonian, 60 Garden Street, Cambridge, MA 02138, USA}

\author[0000-0002-8989-0542]{Kishalay De}
\affil{Department of Astronomy and Columbia Astrophysics Laboratory, Columbia University, 550 W 120th St. MC 5246, New York, NY 10027, USA}
\affil{Center for Computational Astrophysics, Flatiron Institute, 162 5th Ave., New York, NY 10010, USA}

\author{Lynne A. Hillenbrand}
\affil{Cahill Center for Astrophysics, California Institute of Technology, Pasadena, CA 91125, USA}

\author[0000-0001-5926-3911
]{Jill~Juneau}
\affiliation{MIT-Kavli Institute for Astrophysics and Space Research, Massachusetts Institute of Technology, 77 Massachusetts Ave, Cambridge, MA 02139, USA} 

\author[0000-0003-2758-159X]{Viraj R. Karambelkar}
\affil{Cahill Center for Astrophysics, California Institute of Technology, Pasadena, CA 91125, USA}

\author[0000-0002-5619-4938]{Mansi M. Kasliwal}
\affil{Cahill Center for Astrophysics, California Institute of Technology, Pasadena, CA 91125, USA}

\author[0000-0002-4585-9981]{Nathan~P.~Lourie}
\affiliation{MIT-Kavli Institute for Astrophysics and Space Research, Massachusetts Institute of Technology, 77 Massachusetts Ave, Cambridge, MA 02139, USA} 

\author[0000-0001-6331-112X]{Geoffrey~Mo}
\affiliation{Department of Physics, Massachusetts Institute of Technology, 77 Massachusetts
Ave, Cambridge, MA 02139, USA} 
\affiliation{MIT-Kavli Institute for Astrophysics and Space Research, Massachusetts Institute of Technology, 77 Massachusetts Ave, Cambridge, MA 02139, USA} 
\affiliation{MIT LIGO Laboratory, Massachusetts Institute of Technology, Cambridge, MA 02139, USA}

\author[0000-0003-4725-4481]{Sam Rose}
\affil{Cahill Center for Astrophysics, California Institute of Technology, Pasadena, CA 91125, USA}

\author[0000-0003-3769-9559]{Robert A. Simcoe}
\affiliation{Department of Physics, Massachusetts Institute of Technology, 77 Massachusetts
Ave, Cambridge, MA 02139, USA} 
\affiliation{MIT-Kavli Institute for Astrophysics and Space Research, Massachusetts Institute of Technology, 77 Massachusetts Ave, Cambridge, MA 02139, USA}

\author[0000-0003-2434-0387]{Robert D. Stein}
\affil{Department of Astronomy, University of Maryland, College Park, MD 20742, USA}
\affil{Joint Space-Science Institute, University of Maryland, College Park, MD 20742, USA} 
\affil{Astrophysics Science Division, NASA Goddard Space Flight Center, Mail Code 661, Greenbelt, MD 20771, USA}

\begin{abstract}
Episodic mass accretion is the dominant mechanism for mass assembly in the proto-stellar phase. Although prior optical time-domain searches have allowed detailed studies of individual outbursts, these searches remain insensitive to the earliest stages of star formation. In this paper, we present the characterization of two FU Orionis (FUor) outbursts identified using the combination of the ground-based, near-infrared Wide-field Infrared Transient Explorer (WINTER) and the space-based, mid-infrared NEOWISE survey. Supplemented with near-infrared spectroscopic follow-up, we show that both objects are bona fide FUor type outbursts based on i) their proximity to star-forming regions, ii) large amplitude (2-4 magnitudes)  infrared brightening over the last decade, iii) progenitor colors consistent with embedded (Class I) protostars, and iv) ``mixed-temperature'' infrared spectra exhibiting characteristic signatures of cool outer envelopes and a hot inner disk with a wind. While one source, \1, is a known FUor which we independently recover; the second source, \2, is a newly confirmed object. Neither source is detected in contemporaneous ground-based optical imaging, despite flux limits  $\gtrsim 100\times$ fainter than their infrared brightness, demonstrating the capabilities of WINTER to identify heavily obscured young stellar object (YSO) outbursts. We highlight the capabilities of the Galactic Plane survey of the recently commissioned WINTER observatory in addressing the poorly understood FUor population with its unique combination of real-time detection capabilities, multi-color sensitivity, weekly cadence, and wide area coverage. 

\end{abstract}


\section{Introduction} \label{sec:intro}
Young stellar objects (YSOs) grow through episodic accretion, where material from a surrounding disk falls onto the star in bursts rather than at a steady rate \citep[and references therein]{Fischer:2023}. These accretion outbursts significantly increase the system’s luminosity and play a major role in stellar mass assembly \citep{Kenyon:1990, Hartmann:1996, Vorobyov:2005}. Episodic accretion has been proposed as a solution to the ``luminosity problem,'' in which protostars appear dimmer than expected if they were gaining mass at a constant rate throughout their evolution \citep{Kenyon:1990, Dunham:2012}. In addition to influencing how stars gain mass, these bursts can alter the thermal and chemical structure of the disk, affecting processes such as snowline migration and early planet formation \citep{Nayakshin:2023,Lee:2019, 2013ApJ...779L..22J, Kospal:2023}. Discovering and characterizing these outbursts remains essential for understanding their physical origins and role in star formation.

The most extreme form of episodic accretion in YSOs is FU Orionis-type (FUor) outbursts, characterized by large luminosity increases ($\Delta$mag $\gtrsim$ 4–6) and lasting decades or more \citep{Hartmann:1996, Audard:2014, Fischer:2023}. FUor outbursts sustain accretion rates of up to $\gtrsim 10^{-4} \, \text{M}_{\odot} \, \text{yr}^{-1}$, contributing significantly to stellar mass growth in the earliest stages of evolution \citep{Hartmann:1996, Vorobyov:2015}. In contrast, another common category of bursts, EX Lupi-type outbursts (EXors), are less luminous ($\Delta$mag $\gtrsim$ 2–3), shorter-lived (weeks to a few years), and contribute less significantly to the stellar mass budget \citep{Ashraf:2024, EXLup}. Despite the proposed significance of FUors in mass accumulation, they are observationally rare. The classic compilation of \citet{Connelley:2018} reported fewer than 35 FUor-type events had been confirmed using spectroscopic techniques, albeit with most objects having sparse or no photometric coverage of the outburst. 

YSO variability, including FUor outbursts, is commonly studied through time-domain surveys across optical and infrared (IR) wavelengths, often with follow-up spectroscopy for classification. Early discoveries and monitoring of eruptive YSOs relied heavily on optical time-domain surveys, with discoveries in surveys such as ASAS-SN \citep{ASASSN}, PTF \citep{PTF}, and, more recently, Gaia \citep{Gaia17bpi, SE:2020}. However, optical time-domain surveys have limited sensitivity to embedded sources as well as regions of heavy foreground extinction in the Galactic plane and star-forming regions \citep{Enoch:2009, Dunham:2015}.

Recent wide-field IR surveys have proven essential for uncovering outbursts in heavily embedded YSOs, including classes of FUor outbursts previously missed in optical monitoring. The VISTA Variables in the Via Lactea (VVV; \citealt{Minniti:2010}) and its extension VVVX enabled the discovery of dozens of high-amplitude eruptive YSOs, including several FUor candidates in the Galactic plane \citep{Cont:2017, Cont:2017b, Cont:2024, Guo:2024}. Similarly, mid-infrared (MIR) monitoring with WISE \citep{WISE} and NEOWISE \citep{NEOWISE} uncovered a growing population of deeply embedded eruptive variables \citep{Gaia17bpi, Ashraf:2024, Tran:2024, CP:2025}. While the VVV and NEOWISE searches were largely implemented via archival photometric searches together with spectroscopic follow-up many years after the outburst onset, the advent of real-time transient discovery techniques is enabling prompt identification and early spectroscopic follow-up of these outbursts. The ongoing Palomar Gattini-IR (PGIR) survey has uncovered several new NIR-bright FUor events in the northern hemisphere \citep{Hillenbrand:2021, Hillenbrand:2025} and continues to provide a key source of detections, especially toward the Galactic plane (Earley et al., in prep.).

In addition to discovery, multi-color infrared photometry and broad-band spectroscopy play a key role in confirming YSO outbursts and enabling detailed studies of accretion disk evolution. These observations also identify and exclude interlopers, including Mira variables, cataclysmic variables, and classical novae. Multi-wavelength monitoring probes accretion disk formation via changes in NIR and MIR colors---where delays between shorter and longer wavelength bands are critical to understanding the outburst triggering mechanism and instability propagation (outside-in or inside-out; \citealt{Bell:1995, Cleaver:2023, Nayakshin:2024}). Simultaneously, such coverage traces the development of hot inner disk regions, the clearing or rebuilding of circumstellar material and the presence of infalling envelopes or outflows \citep{Hillenbrand:2022, Caratti:2017}. \cite{Connelley:2018} demonstrate that NIR spectroscopy provides a robust basis for classifying FUor-type objects, which exhibit a distinctive combination of strong molecular absorption, weak or absent emission lines, low-gravity spectral features, and a mixed-temperature spectrum that becomes progressively cooler at longer wavelengths.

\begin{figure*}
\centering
\includegraphics[width=0.5\textwidth]{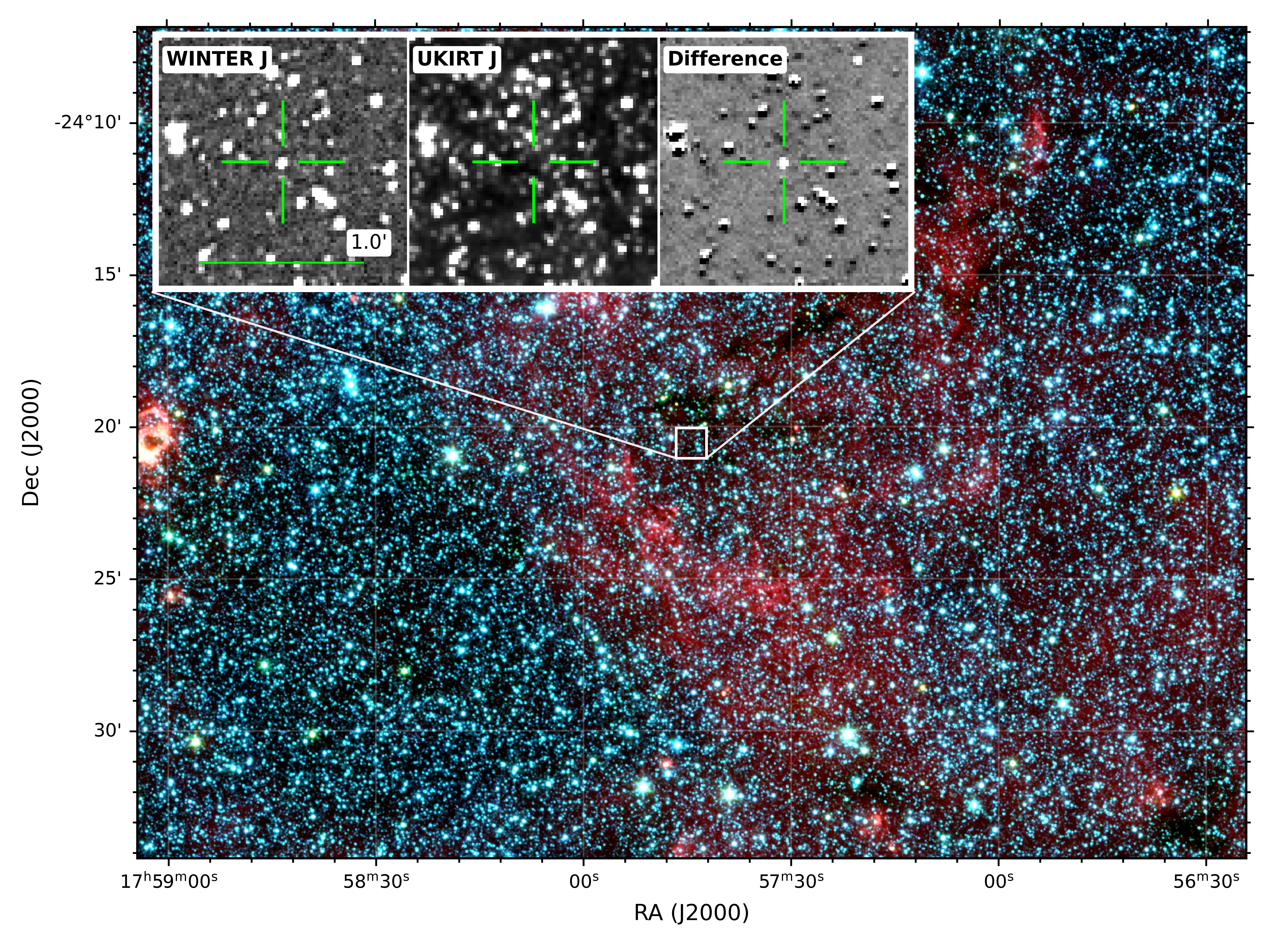}%
\hspace{0.02\textwidth}%
\includegraphics[width=0.47\textwidth]{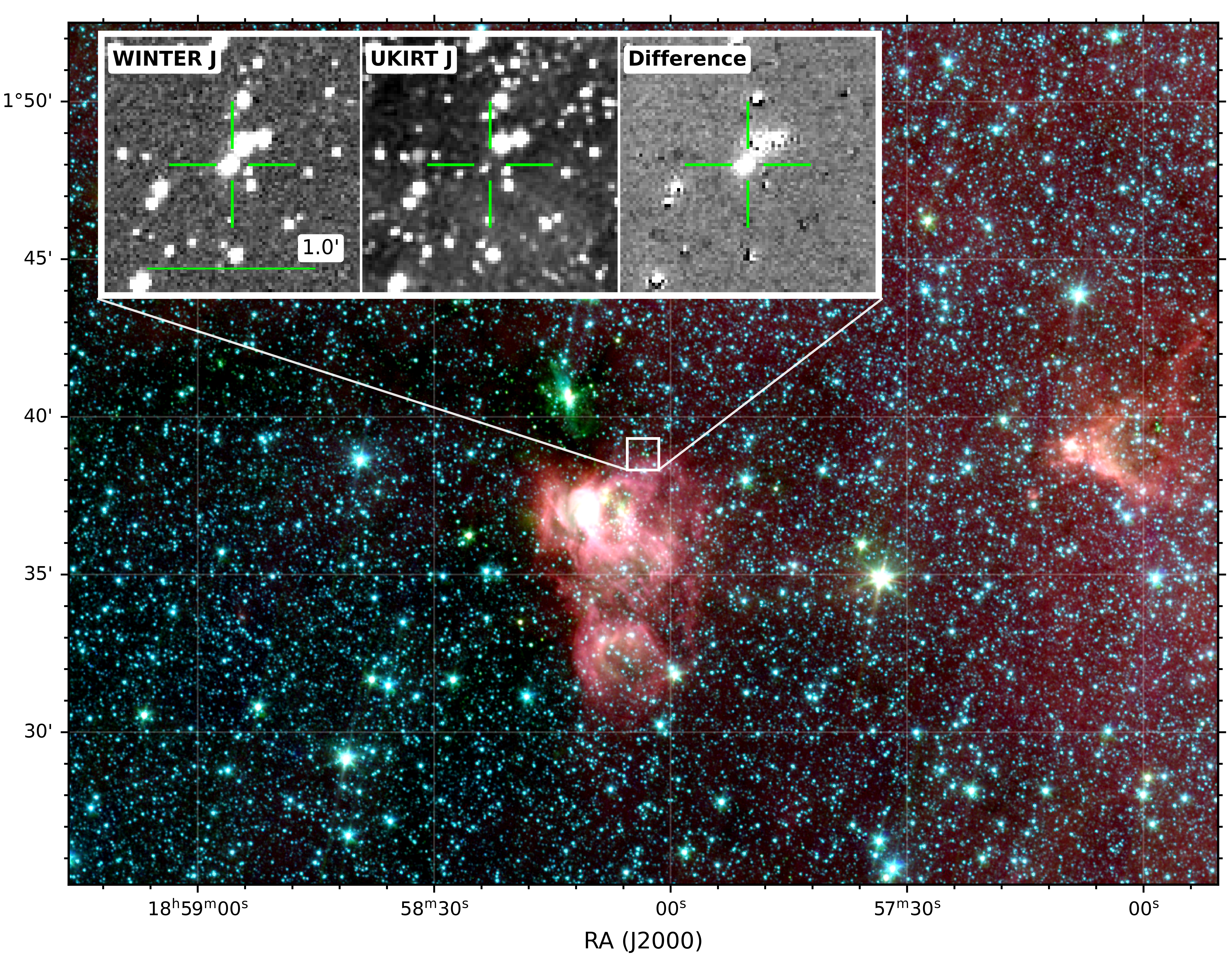}
\caption{WINTER $J$-band detections of \1\ (left) and \2\ (right) overlaid on images of the events' environments. The larger image shows the area surrounding the outburst taken with the Spitzer Space Telescope \citep{Spitzer} with IRAC channels 1 (3.6 $\mu$m), 2 (4.5 $\mu$m), and 4 (8.0 $\mu$m) used as blue, green, and red, respectively, in the composite images. Inset: details from WINTER science images showing the initial detection of the event (left), a corresponding archival reference image from UKIRT (middle), and the difference image between WINTER and UKIRT (right).}
\label{fig:WNTR24_cutouts}
\end{figure*}

The Wide-Field Infrared Transient Explorer (WINTER) is a new NIR time-domain survey on a dedicated 1-meter robotic telescope at Palomar Observatory \citep{Lourie:2020, Frostig:2024, Frostig:2025}. Each pointing covers a 1.2 $\text{deg}^2$ field of view with a 1.1" pixel scale and the option of observing in Y, J, or a shortened H (Hs) filters, centered on 1.02, 1.25, and 1.60 $\mu m$. WINTER covers the Galactic plane during the summer months, running multiple surveys at an approximately two-week cadence. Unlike VVV and PGIR which almost entirely operated their time-domain surveys in a single filter (in Ks and J band, respectively), WINTER carries out multi-color cadenced surveys of the Galactic plane. Like PGIR, WINTER also includes the infrastructure for rapid discovery of transients with a real-time image subtraction pipeline, operating with smaller field of view (1.2 $\text{deg}^2$ vs. 25 $\text{deg}^2$) but substantially higher depth (median $J_{\mathrm{lim}} = 18.5$ vs. $15$ AB mag) and spatial resolution (1.1\arcsec vs 8.7\arcsec) than PGIR \citep{De:2020}. 

In this study, we demonstrate WINTER's potential for discovery and characterization of YSO outbursts via the detection and characterization of heavily obscured FUor outbursts identified independently via searches in NEOWISE. We present two FUors studied in NIR with WINTER, in MIR with WISE and NEOWISE, and with NIR spectroscopic follow up. One source, \1 is a known YSO outburst, also known as 2017fzn and as L222\_192 in \cite{Guo:2024}, which also classified the object as an FUor. The second source, \2, is a newly characterized FUor, adding the expanding sample of optically obscured but IR-bright outbursts. Of note, \2 was recently proposed as a candidate FUor by \citet{CP:2025}, based on a long-term rise in NEOWISE photometry, although it has not been confirmed spectroscopically before this study.

We present the observations of \1 and \2 in WINTER, WISE, and other surveys, the source environment, and spectroscopic follow-up in Section \ref{sec:obs}. In Section \ref{sec:analysis}, we analyze the observations, presenting a study of the host environments, the progenitors of the outbursts, and spectroscopically classify the FUors. We discuss the implications for future observations of YSO outbursts with WINTER and conclude in Section \ref{sec:disc}.

\begin{figure*}
    \centering
    \includegraphics[width=\textwidth]{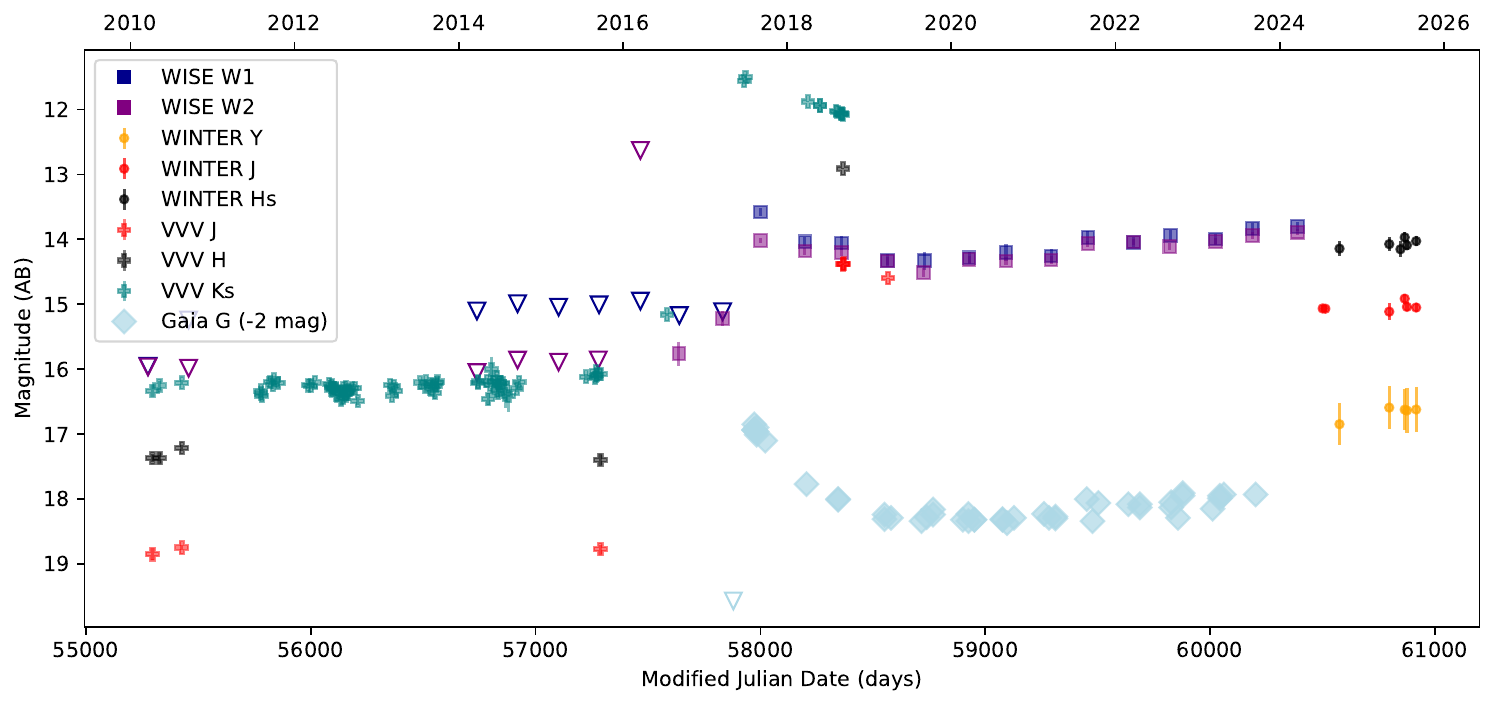}


    \includegraphics[width=\textwidth]{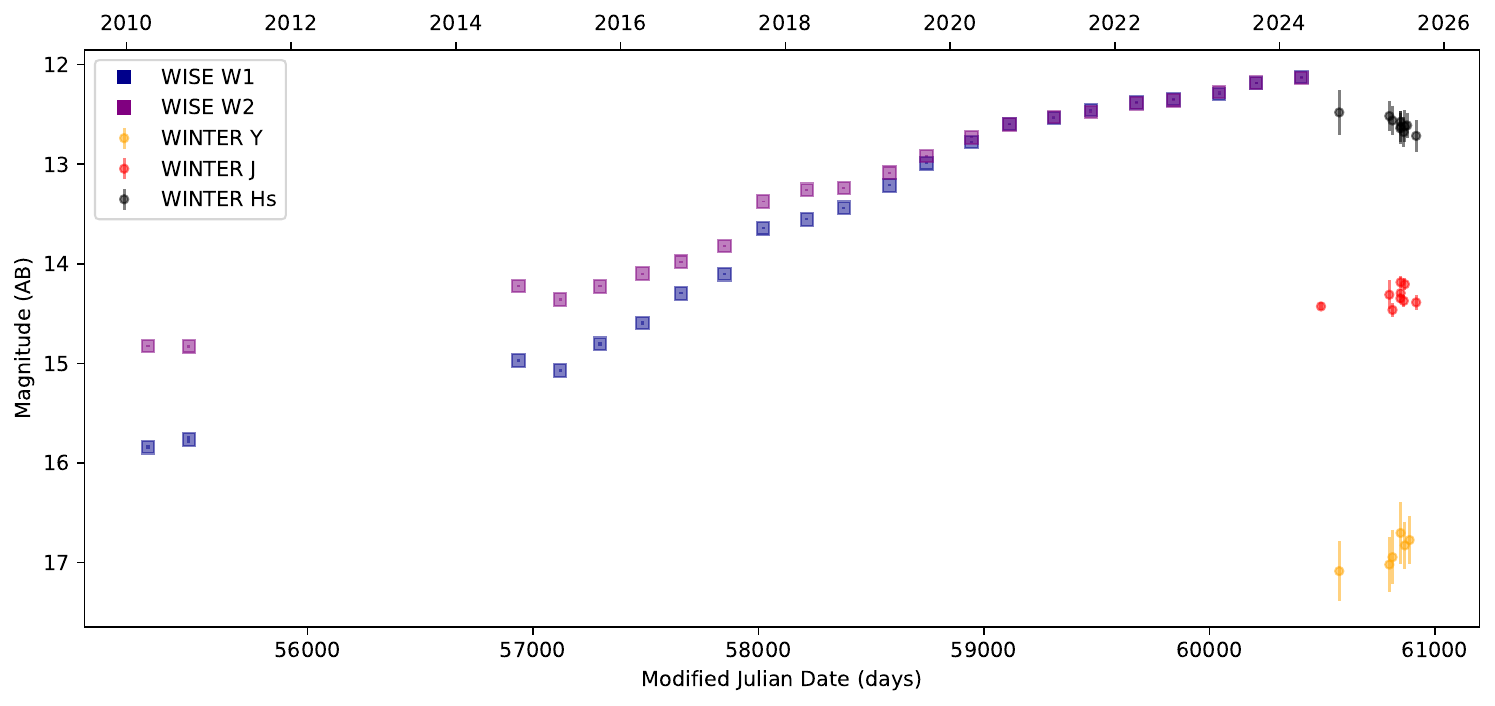}

    \caption{Multi-band light curves for \1 (top) and \2 (bottom) combining data from WISE/NEOWISE ($W1$, $W2$), WINTER ($Y$, $J$, $H_s$), VVV ($J$, $H$, $K_s$), and Gaia ($G$). Inverted open triangles mark 5$\sigma$ upper limits corresponding to non-detections. For \1, the progenitor is detected in VVV before the outburst but not in WISE/NEOWISE. The outburst begins in 2016, visible in VVV $K_s$ and WISE $W2$. The initial rise is not detected in WISE $W1$ or Gaia $G$, which provide upper limits at that time. The peak and decline are captured from Gaia $G$ through WISE $W2$, with WINTER observations starting as Gaia and WISE coverage end. In contrast, \2 shows a slower outburst beginning in 2014 that continues through 2025, covered by WISE and WINTER. Neither source is detected in the optical by ZTF to $\sim$20 mag, confirming that both are heavily obscured at optical wavelengths. For clarity, Gaia $G$ has been shifted upward by 2 magnitudes, highlighting how much fainter the optical detections are compared to the infrared bands.}
    \label{fig:WNTR24_lc}
\end{figure*}

\section{Observations} \label{sec:obs}

The two FUor events presented here were first identified as mid-infrared transients in systematic searches of NEOWISE data. As part of WINTER’s broader Galactic plane survey, we also targeted regions with NEOWISE outbursts to verify near-infrared brightness and obtain multi-color light curves. In this paper, we present detailed studies of two objects as a pilot demonstration of WINTER’s capabilities for characterizing obscured transients in the Galactic plane.

\subsection{Discovery, NEOWISE, and WINTER initial observations} \label{sec:winter}

Systematic processing of NEOWISE data with the unWISE ZOGY pipeline \citep{Lang:2014, Meisner:2018, De:2023} flagged both sources as mid-infrared transients. The pipeline performs image subtraction between the original WISE coadds from 2010 and time-resolved NEOWISE coadds obtained from 2014–2024, producing a catalog of statistically significant transient candidates. The search identified both events as transients and assigned database names as part of the WISE Transient Pipeline (WTP). \1 and \2 are located at J2000 coordinates $\alpha = 269.435136$, $\delta =  -24.342384$ and $\alpha = 284.514614$, $\delta = +1.646963$ and assigned WTP designations WTP20aabowt and WTP14acbxtp, respectively.

WINTER first observed both FUors in the $J$ band on UT 2024-07-10 (\1) and 2024-07-03 (\2) as part of the Galactic plane program. The data were processed with the WINTER pipeline implemented with mirar (Karambelkar and Stein et al., in prep), an open-source, modular \texttt{python}-based system. The pipeline runs all reduction steps including calibration, astrometry, stacking, difference imaging, and transient identification. Difference images were generated using reference images from UKIRT for $J$ and $Hs$, and from Pan-STARRS $Y$, where available. Figure~\ref{fig:WNTR24_cutouts} shows example image cutouts produced by the pipeline from the original WINTER $J$-band images, the UKIRT reference images, and the resulting difference images for both events. \1 and \2  were flagged as candidate transients and assigned WINTER database names, WNTR24gxcua and WNTR24fxegv. For readability throughout this work, we refer to them using shortened forms of these names.

The sources were selected for WINTER follow-up based on their high near-infrared brightness and coincidence with previously identified WISE transient candidates. Monitoring has continued through the 2025 Galactic plane season in the $Y$-, $J$-, and $Hs$-bands as part of WINTER’s ongoing infrared Galactic plane program. 

\subsection{Optical Observations} \label{sec:Gaia}

There are no optical detections of these IR-bright outbursts, with the exception of faint Gaia \citep{Gaia} $G$-band coverage for \1. The Gaia $G$-band has broad coverage from $\sim 0.3 - 1.0 \mu m$\footnote{https://www.cosmos.esa.int/web/gaia/iow\_20180316} , including some NIR flux around 1 $\mu m$. The Gaia data were downloaded from the Gaia Photometric Science Alerts webpage\footnote{https://gsaweb.ast.cam.ac.uk/alerts/}, where the event is designated as Gaia 17bzk. These data include $G$-band flux but no error bars. The Gaia alert data includes non-detections before the event and the decline of the outburst between UT 2017-08-07 and UT 2023-09-16 (as seen in Figure \ref{fig:WNTR24_lc}). 

Archival data from the Zwicky Transient Facility \citep[ZTF]{ZTF}, as accessed through the ZTF forced photometry server, revealed no detection of the outburst in available data between UT 2018-04-04 and UT 2024-12-20 in the more narrow $g$, $r$, or $i$ filters to an average 5$\sigma$ limiting depth  of $\sim$20 mag AB in each band. This implies these sources are quite obscured at optical wavelengths, though bright at NIR and MIR wavelengths.

\subsection{Light curve evolution} \label{sec:wise}

Both events have long-term infrared monitoring from the WISE and NEOWISE missions (Figure~\ref{fig:WNTR24_lc}). They were first observed in two epochs during the original WISE mission in 2010, with continued coverage in the $W1$ and $W2$ bands by NEOWISE at a six-month cadence from 2014 through the final visit in 2024. 

\1 was not detected in WISE until early 2016, when it began a sharp rise in the $W2$ band, brightening by 1.7 magnitudes over 364 days (Figure~\ref{fig:WNTR24_lc}). This was followed by a decline of 0.5 magnitudes in 2019, after which the source exhibited a more gradual increase of 0.62 magnitudes between 2020 and 2024. The initial rise was not seen in $W1$, likely due to the band’s $\sim$1 magnitude shallower depth; however, near peak brightness the source was detected at $W1 = 13.6$ mag, with a color of $W1 - W2 = 0.4$ mag. By 2024, the emission became bluer with $W1 - W2 = 0.1$ mag. The brightening is also evident in VVV $K_s$, which shows the onset of the outburst in 2016, with the progenitor detected prior to the rise in $J$, $H$, and $K_s$ (data from \citealt{Guo:2024}). In contrast, Gaia $G$-band observations do not detect the source up through May 2017, but then observe the outburst three months later, implying a rapid optical rise after the start of the NIR and MIR outburst.

\2, in contrast, has been consistently detected in WISE since 2010. The earliest data show no significant variability, but a brightening of 0.9 (0.6) magnitudes in $W1$ ($W2$) began in 2014, following a brief dip, and continued steadily to an overall change of 3.7 (2.7) magnitudes on in $W1$ ($W2$) by 2024. The $W1 - W2$ color has decreased throughout this evolution, beginning at 1.0 magnitudes in 2010 and trending toward a nearly monochromatic value by 2020.

WINTER observations continue to follow both sources through the 2024 and 2025 seasons, with both sources showing a continuous, gradual brightening in all three filters. 

\subsection{Archival imaging of the source environment}

Both sources were observed by {\it Spitzer}/IRAC in all four channels as part of the GLIMPSE survey \citep{Benjamin:2003}. We retrieved the data to provide both progenitor photometry and environmental context. Point-source photometry was obtained from the {\it Spitzer} source catalog \citep{GLIMPSE}, and wide-field color images were generated using archival IRAC mosaics from the CDS/P/SPITZER/color HiPS survey, accessed via the hips2fits service. We used the Aitoff projection centered on each FUor’s coordinates, with a field of view of 0.6 degrees and an output resolution of 1024 x 1024 pixels; the resulting FITS cubes were split into individual R, G, and B channels corresponding to the 8.0, 4.5, and 3.6 $\mu$m IRAC bands, and combined into the composite image shown in Figure \ref{fig:WNTR24_cutouts}.

\subsection{Spectroscopic observations} \label{sec:spectra}

\begin{figure*}
    \centering
    \includegraphics[width=\linewidth]{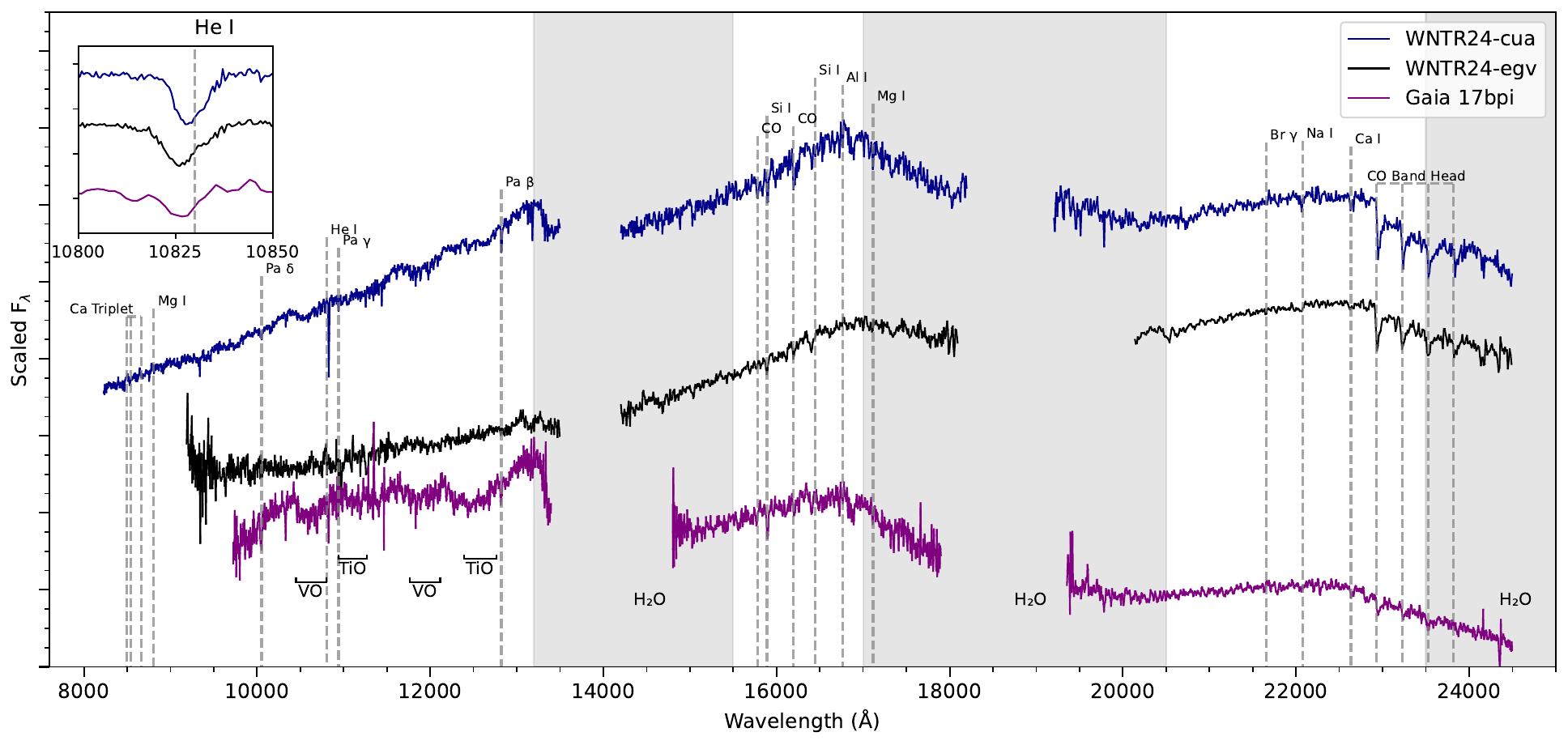}
    \caption{Spectra for the FUors compared with Gaia 17bpi \citep{Gaia17bpi} in purple. The inset highlights the He I $\lambda$10830 line demonstrating blue-shifted absorption, indicative of a strong wind. Blue: \1 spectrum taken with FIRE. Black: \2 spectrum taken with SpeX with an inset of the He I line from a FIRE spectrum. High noise regions are masked. All three spectra show mixed temperatures typical of FUors, with deep molecular absorption at longer wavelengths, notably the CO bandhead near 2.3 $\mu$m, which becomes less prominent at shorter wavelengths.
   } 
    \label{fig:spectra}
\end{figure*}

We obtained NIR spectroscopy of the sources with the SpeX spectrograph on the NASA Infrared Telescope Facility \citep{SpeX} and the Folded-port Infrared Echellette \citep[FIRE]{FIRE} on the Magellan telescopes (Figure \ref{fig:spectra}). \1 was observed with FIRE on UT 2024-07-04, while \2 was observed with SpeX on UT 2022-10-08 and with FIRE on UT 2024-07-11. The SpeX data were taken in the SXD mode with a 0.8\arcsec slit, for a total exposure time of 30 minutes as part of program 2022B059 (PI: De). The data were reduced with the \texttt{spextool} pipeline \citep{Cushing:2004} while telluric and flux calibration were carried out using the \texttt{xtellcor} tool \citep{Vacca:2003}. The FIRE data were taken in the prism cross-dispersed echelle mode with a 0.6" slit, which delivers spectra at R=6000 across $0.82–2.51 \, \mu$m, and reduced with the \texttt{PypeIt} package \citep{pypeit:joss_arXiv}.

\section{Analysis}
\label{sec:analysis}

\subsection{Environment} \label{sec:env}

Both \1 and \2 lie next to nebulosity seen in the Spitzer images (Figure \ref{fig:WNTR24_cutouts}). \1 is in the middle of some nebulosity but no obvious large HII region. \citet{Guo:2024} report a projected association between \1 and a nearby star-forming region located at a heliocentric distance of 6.1~kpc ($d_{\mathrm{SFR}}$), based on the presence of a star-forming region within 5~arcmin of the source. They find, however, a significantly smaller kinematic distance of $3.0^{+1.1}_{-2.0}$~kpc ($d_k$), derived from CO absorption line radial velocities. These values are not consistent within uncertainties, suggesting that \1 likely lies in the foreground of the associated SFR. \citet{Guo:2024} caution that $d_{\mathrm{SFR}}$ is only a rough estimate and may be affected by projection effects, while $d_k$ may be biased by unresolved binarity or peculiar motion. The offset in distance implies that \1 may be forming in a more isolated or diffuse environment rather than within a clustered star-forming region.

\2 lies in close proximity to the H\,\textsc{ii} region [DLW84] G35.2S, with G35.2N also nearby. Using \textit{Gaia} EDR3 parallax measurements (\( \varpi = 0.397 \pm 0.025 \)~mas; \citealt{gaiadr3}), we estimate a distance of \( 2.52 \pm 0.16 \)~kpc to the region. The projected angular separation between \2 and the center of G35.2S is \( 158.3'' \), corresponding to a physical offset of only \( 1.93 \pm 0.12 \)~pc at this distance, indicating a likely association. 
A dense molecular core (W48 89) identified by \cite{keown2019} is essentially identical in position with \2.
Many sources in the immediate vicinity were identified by \cite{spicy} as YSOs, although not the progenitor of \2. 

Following the procedure outlined in \citealt{Tran:2024}, we also estimate the distance to the outbursts by identifying nearby candidate YSOs with known distances. First, we take sources within 20' of the FUors from the Spitzer (SEIP) source list \citep[SSTSL2,][]{2021yCat.2368....0S} and the 2MASS All Sky Catalog of point sources \citep{2003tmc..book.....C} and apply a color cut of $J \text{-} K \textgreater 0.8 $ and $[3.6] \text{-} [4.5] \textgreater 0.15$ to identify sources with IR-color excess as likely YSOs. IR-excess contaminants, such as star-forming galaxies, are eliminated following the procedures outlined in \cite{Gutermuth:2009}. 

This results in 466 and 273 candidate YSOs for \1 and \2 respectively. For \1, 17 of the candidate YSOs have counterparts in the Gaia DR3 catalog \citep{gaiadr3} with measured parallaxes, yielding a median distance of $2.4^{+0.7}_{-0.7}$ kpc. Due to the discrepancies in the estimated distances presented in \citealt{Guo:2024}, we use the estimate of 2.4 kpc for the analyses in Section \ref{sec:sed}. Similarly, for \2, 10 of the candidate YSOs have Gaia distances ranging from 0.2 to 2.2 kpc. However, their relatively uniform spatial distribution offers no clear basis for a reliable distance estimate, and many of the stars in this region are likely too obscured for Gaia detections. We therefore adopt the distance of the nearby H\,\textsc{ii} region at 2.52 kpc.

\subsection{Spectroscopic features and classification}

Both \1 and \2 are spectroscopically similar to other FUors at NIR wavelengths \citep[as explored in detail in][]{Connelley:2018}. The spectra are shown in Figure \ref{fig:spectra} with another FUor, Gaia17bpi \citep{Gaia17bpi}, for comparison. All three spectra demonstrate a ``mixed-temperature'' spectrum characteristic of FUors. At the shortest infrared wavelengths (around 1 $\mu$m), the spectra resemble a late-F to early-G type star at T$\sim$6000 K. However, at longer wavelengths the spectra are consistent with an M-type star at T$\sim$3000 K with deep molecular absorption features. For example, there is prominent absorption from the $^{12}$CO band-head (around $2.3 \mu m$) and H$_2$O band ($1.3- 1.5 \mu$m and $1.75 - 2.05\mu$m). Molecular absorption from VO and TiO is strongest in Gaia17bpi, present but weaker in \1, and not confidently detected in \2.  Additionally, there are atomic lines from Al I, Ca I, Mg I, Na I, and Si I, along with H Paschen lines in both objects. 

Both of the outbursts show strong He I $\lambda$10830 blue-shifted absorption, which is evidence for a strong wind \citep{Connelley:2018}. We estimate a FWHM velocity of approximately 231 km/s and 305 km/s for \1 and \2, respectively, and corrected for instrumental broadening. Measuring against the blue end of the absorption feature, we estimate a maximum velocity of 221~km/s and 321~km/s, for \1 and \2.

\begin{figure}
    \centering
    \includegraphics[width=1\linewidth]{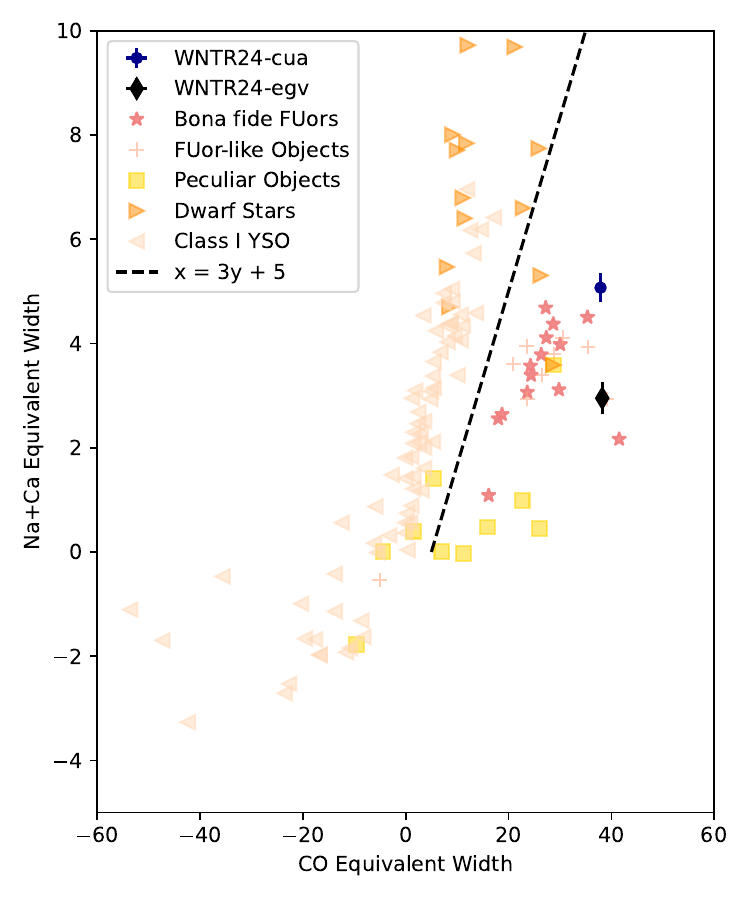}
    \caption{Classification scheme and data adapted from \cite{Connelley:2018} with the two WINTER sources over plotted. The WINTER sources fall near other bona fide FUors and show notably high CO equivalent widths. }
    \label{fig:classification_yso}
\end{figure}

Additionally, both WINTER sources exhibit steep red continua across the near-infrared, consistent with heavily reddened, embedded systems. As shown in Figure~\ref{fig:classification_yso}, their spectra are significantly redder than Gaia~17bpi, suggesting a greater degree of circumstellar extinction and/or cooler outer disk temperatures. The continuum shape in both cases rises sharply from the $J$ to $K$ bands, with minimal flux at shorter wavelengths. This behavior is consistent with their non-detections in ground-based optical surveys: archival ZTF data show no detections in $g$ or $r$ bands down to a 5$\sigma$ limit of $\sim$20 mag, indicating that these outbursts are effectively invisible at optical wavelengths. These limits imply $g-J \gtrsim 5$ mag, consistent with their red continua. The detection of \1 in Gaia is likely due to the broad Gaia $G$ bandpass, which extends red-ward to nearly 1~$\mu$m and can capture flux in the far-red continuum where these objects still emit. The steep spectral slopes, in combination with their strong molecular absorption features, support the classification of these objects as deeply embedded FUor-type outbursts.

Based on these spectroscopic properties, we classify \1 and \2 as FUors following the criteria outlined in \cite{Connelley:2018}. We measure the equivalent widths of CO, Na I, and Ca I, as outlined in \citet{Messineo:2021}, and compare to bona fide FUors from the \cite{Connelley:2018} sample (Figure \ref{fig:classification_yso}). Both \1 and \2 fall within the bona fide sample, with notably large CO equivalent widths. Additionally, both sources exhibit the eight NIR spectral characteristics of FUors outlined in \cite{Connelley:2018}. These characteristics include CO absorption and molecular band absorption, metallic, hydrogen, and He I lines, and changing of spectral type with wavelength. 

\subsection{Spectral energy distribution}  \label{sec:sed}

\begin{figure*}
\centering
\includegraphics[width=0.48\textwidth]{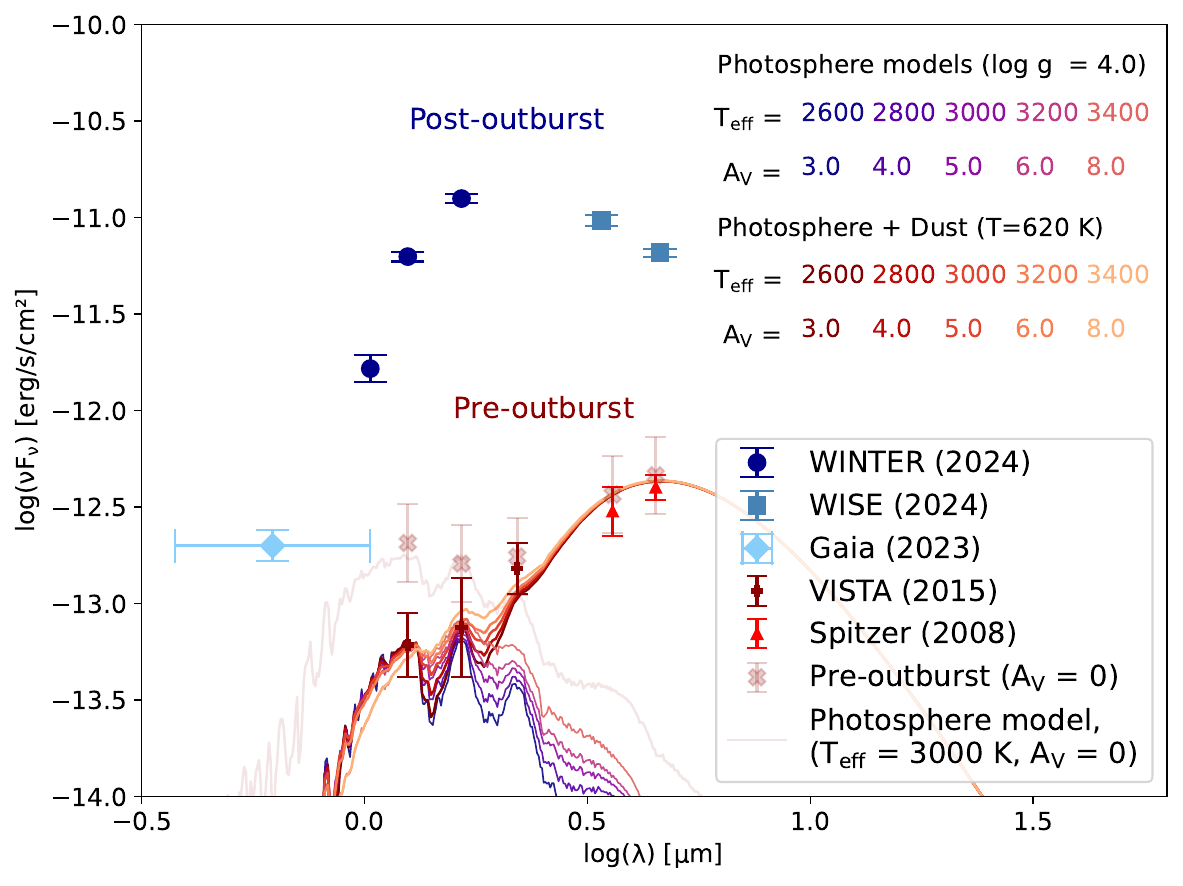}%
\hfill%
\includegraphics[width=0.48\textwidth]{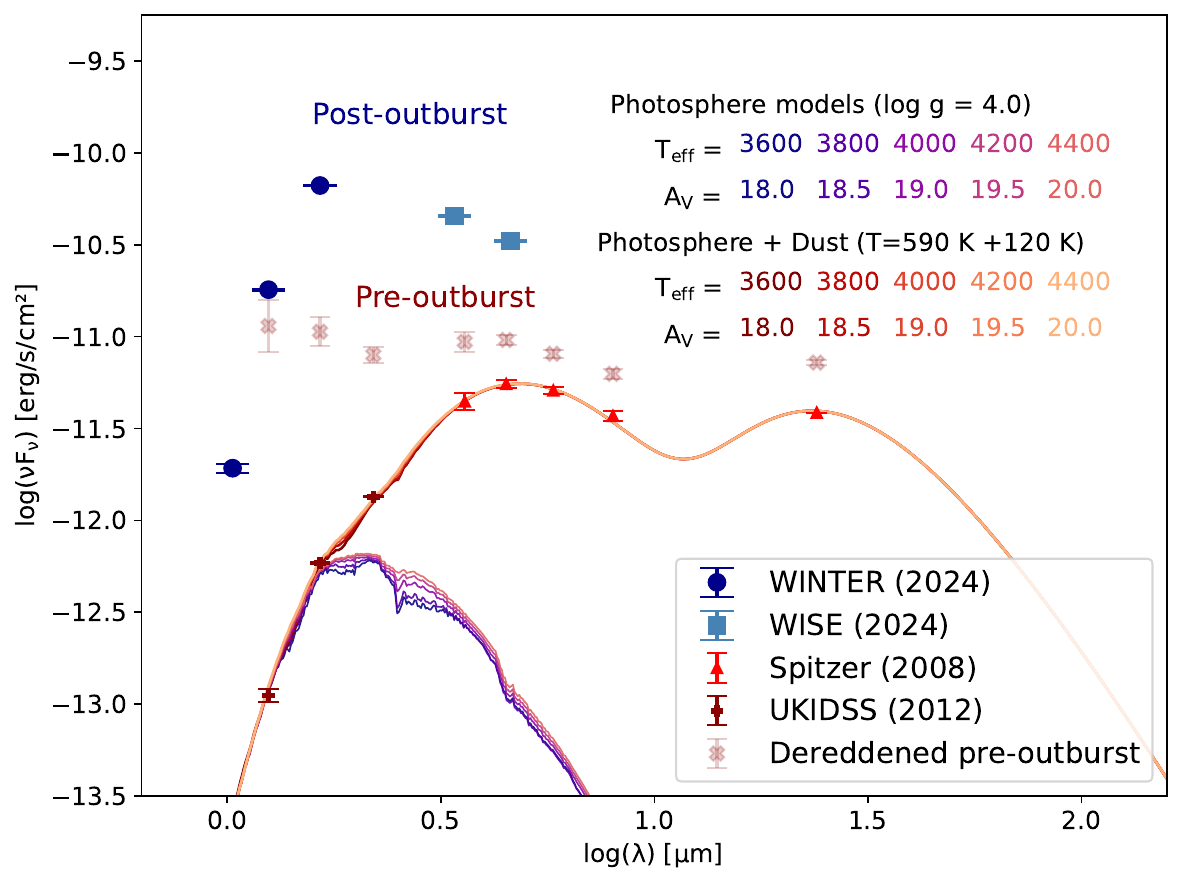}
\caption{The spectral energy distributions (SEDs) for \1 (left) and \2 (right) showing pre- (reds) and post-outburst (blues) energy. The pre-outburst systems are shown with reddened photospheric models with a range of temperatures and extinctions in purple lines. The same models with additional blackbody components due to dust are shown in orange lines to explain the IR excess. The dereddened ($A_V = 0$) pre-outburst data are shown, along with a representative dereddened photosphere model, using extinction corrections of $A_V = 5.0$~mag and $A_V = 19.0$~mag, for \1 and \2, respectively, derived from the photospheric model fits. }
\label{fig:WNTR24_seds}
\end{figure*}

Archival infrared imaging reveals that both \1 and \2 have luminous progenitors, enabling reconstruction of their pre- and post-outburst spectral energy distributions (SEDs; Figure~\ref{fig:WNTR24_seds}) and investigating both the pre-main-sequence nature of the progenitors and the total luminosity increase during outburst. \1 is detected in deep near- and mid-infrared surveys, including VISTA VVV Data Release 2 \citep{VVV2}, the Spitzer GLIMPSE Source Catalog (I, II, and 3D; \citealt{GLIMPSE}), and DECaPS DR2. \2 lacks optical detections but appears in 2MASS (source ID 18580350+0138499), Spitzer/GLIMPSE, the UKIDSS Galactic Plane Survey \citep{UKIDSS}, and WISE. \1 was not reported in The Spitzer/IRAC Candidate YSO Catalog for the Inner Galactic Midplane (SPICY) catalog because the progenitor is too faint \citep{spicy}, while \2 does have a listed progenitor in the catalog.

Figure~\ref{fig:WNTR24_seds} presents SEDs spanning the optical to mid-infrared for both sources before and after their outbursts. The pre-outburst SED for \1 combines 2008 Spitzer and 2015 VISTA photometry, while the post-outburst data are from 2023 Gaia and 2024 WINTER and WISE data. For \2, pre-outburst data come from 2008 Spitzer and 2012 UKIDSS observations, with post-outburst photometry from 2024. Although these epochs are not contemporaneous, they represent comparable phases in each source’s evolution and are useful for characterizing the progenitor system.

We model the progenitors of each outburst with a combination of photospheric models \citep[from][]{nextgen} and black bodies approximating dust in the system. We fit the photospheric models to the $J$- and $H$-band data with a surface gravity of log(g)=4.0---typical for an M dwarf star---and a range of temperatures and extinctions. The addition of black body SEDs model the NIR and MIR excess which could result from an extended dust envelope around the progenitor. For \1, we find a range of models with $T_{\text{eff}} = 2600 - 3400$ K at extinctions of $A_{V} = 3.0 - 8.0$ reasonably fit the data. The addition of a $T = 620$ K black body helps model the MIR emission.

\2 is similarly well modeled with a cool star, though slightly hotter and with higher extinction. The models span $T_{\text{eff}} = 3600 - 4400$ K with $A_{V} = 18.5 - 20.5$. For dust in \2, we use a two-component black body model with $T = 590$ and $120$ K. 

To place the SED-based extinction estimates in context, we also derive values from spectral dereddening and Galactic dust maps. First, we estimate the visual extinction ($A_V$) toward each source by dereddening their near-infrared spectra using the \citet{CCM89} extinction law ($R_V = 3.1$) until their continuum slopes match that of Gaia\,17bpi, which has a known extinction of $A_V \approx 3.5$\,mag. This comparison yields relative extinctions of $\Delta A_V \approx 1.6$\,mag and $10.0$\,mag, corresponding to values of $A_V \approx 5.1$\,mag for \1 and $A_V \approx 13.5$\,mag for \2. We then compared these values to predictions from the Bayestar 3D Galactic dust map \citep{Green:2019}. For \1 (2.4 kpc), the cumulative Galactic extinction is $A_V \approx 4.3$ mag, rising to only $A_V \approx 4.9$ mag along the full line of sight, indicating that the source already lies behind most of the Galactic dust in this direction. In contrast, \2 (2.52 kpc) shows $A_V \approx 3.8$ mag out to its distance compared to $A_V \approx 8.4$ mag for the entire column, implying that only part of the Galactic dust lies in the foreground and a substantial fraction remains behind the source. The much higher extinction estimate from the SED fit ($A_V \approx 19$ mag) therefore indicates significant local extinction in the circumstellar environment. For the analysis in Section~\ref{sec:mdot}, we adopt the extinction estimates from the SED fits.

The SEDs of the progenitor systems also enable classification of the proto-stars via their spectral indices. For data between 2.2 and 10 $\mu$m, we get spectral indices of $\alpha = +1.36, +0.80$ for \1 and \2, respectively. Following the classification scheme in \citet{Greene:1994}, both sources are Class I YSOs. In other classification schemes extending to longer wavelengths, the progenitor of \1 could be considered a ``flat spectrum'' type YSO, similar to the FUor discussed in \cite{Tran:2024}. These spectral indices together with the dust excess and high extinction indicate the objects are Class I, although SPICY reports \2 as a Class II object based on [4.5] - [8.0] colors \citep{spicy}.

Figure 5 also presents the dereddened pre-outburst SEDs assuming $A_V = 5.0$ mag for WNTR24-cua and $A_V = 19.0$ mag for WNTR24-egv (following extinction curves from \citealt{Gordon:2023}, which builds upon \citealt{Gordon:2021, Decleir:2022}). These values represent estimates of the total foreground extinction, which likely includes both (mostly) interstellar and circumstellar components that are difficult to separate. The figure also presents one dereddened, representative photospheric model to show a simplified estimate of the intrinsic luminosity of just the pre-outburst object without the dust modeled by black bodies. 

\subsection{Luminosities and mass accretion rates} \label{sec:mdot}

We estimate the mass accretion rates of both FUor outbursts by combining physical parameters derived from pre-main-sequence stellar models with bolometric luminosities obtained from the SEDs. For a simplified estimate of the total luminosity, we integrate the extinction-corrected SED between 1 and 30~$\mu$m for both the pre- and post-outburst luminosity ($L_{\rm pre}$ and $L_{\rm post}$), which includes stellar, disk, and envelope emission. This provides a consistent measure of the systems' total radiative output, including both reprocessed and accretion-powered components, though it ignores the potentially complex geometry of the system and likely overestimates the intrinsic luminosity of the central source due to the inclusion of reprocessed emission from the surrounding envelope. To isolate just the luminosity from the proto-star ($L_{\star}$), we integrate the extinction-corrected ($A_V = 0$) photosphere model to isolate the central source from the surrounding dust. As a note, the post-outburst photometry used in this analysis was obtained in 2024, years after the outburst onset for each event. 

For \1, we adopt $A_V = 5.0$~mag and find a total pre-outburst luminosity of $L_{\rm pre} \approx 0.6\,L_\odot$, and post-outburst luminosity of $L_{\rm post} \approx 5.5\,L_\odot$. In \2, where $A_V = 19.0$~mag, we estimate $L_{\rm pre} \approx 5.6\,L_\odot$ and $L_{\rm post} \approx 222.1\,L_\odot$. 

To estimate the stellar mass and radius of the progenitors, we interpolate the 1~Myr pre-main-sequence evolutionary models of \citet{Baraffe:2015} at the effective temperature and pre-outburst bolometric luminosity of each system. These luminosities include contributions from both the stellar photosphere and circumstellar dust re-emission. Thus, the inferred stellar masses and radii should be regarded as upper limits; the true stellar values may be lower if a significant fraction of the bolometric luminosity arises from non-stellar components. For \1, adopting $T_{\rm eff} = 3000$~K and $L_{\star} = 0.033\,L_\odot$, we infer $M_* = 0.15\,M_\odot$ and $R_* = 0.86\,R_\odot$. For \2, with $T_{\rm eff} = 4000$~K and $L_{\rm pre} = 2.97\,L_\odot$, we find $M_* = 0.60\,M_\odot$ and $R_* = 2.00\,R_\odot$.

Assuming the excess luminosity is powered by accretion from a circumstellar disk, we use the standard expression for accretion luminosity,
\[
L_{\rm acc} = \frac{G M_* \dot{M}}{R_*} \left(1 - \frac{R_*}{R_{\rm in}} \right),
\]
with $R_{\rm in} = 5\,R_*$ as the adopted inner disk truncation radius \citep{Gullbring:1998}. Applying this relation, we find that $L_{\rm post} = 5.5\,L_\odot$ in \1, measured approximately 7~years after the onset of its outburst, corresponds to an accretion rate of $\dot{M} \approx 2.1 \times 10^{-6}\,M_\odot\,\mathrm{yr}^{-1}$. For \2, the larger increase of $L_{\rm post} = 222.1\,L_\odot$, derived from observations taken roughly 10~years after the outburst began, implies $\dot{M} \approx 3.0 \times 10^{-5}\,M_\odot\,\mathrm{yr}^{-1}$. These derived values are summarized in Table \ref{tab:wntr_params}.

These estimates fall within the expected range for FUor outbursts. Typical accretion rates of $\sim10^{-6}$--$10^{-4}\,M_\odot\,\mathrm{yr}^{-1}$ have been reported both in observational surveys \citep[e.g.,][]{Audard:2014} and recent theoretical studies \citep[e.g.,][]{Nayakshin:2024b}, placing \1 at the lower end of the distribution and \2 near the median. The modest accretion rate inferred for \1 is consistent with a low-mass progenitor and is similar to that of Gaia\,17bpi, a low-luminosity FUor with comparable properties.

Given the uncertainty in the distance to \1, which ranges from $d = 2.4$~kpc (calculated in Section~\ref{sec:env}) to the two proposed values of $d = 3.0$ and $6.1$~kpc from \citet{Guo:2024}, we scale the luminosities and accretion rates accordingly. At 3.0~kpc, the post-outburst luminosity increases to $L_{\rm post} \approx 8.6\,L_\odot$ and the inferred accretion rate to $\dot{M} \approx 2.0 \times 10^{-6}\,M_\odot\,\mathrm{yr}^{-1}$. At 6.1~kpc, these values become $L_{\rm post} \approx 35.6\,L_\odot$ and $\dot{M} \approx 8.4 \times 10^{-6}\,M_\odot\,\mathrm{yr}^{-1}$. These values remain within the expected range for FUors but suggest that \1 may be more luminous than implied under the nominal 2.4~kpc assumption. 

\section{Discussion and Summary} \label{sec:disc}

\begin{figure*}
    \centering
    \includegraphics[width=1\linewidth]{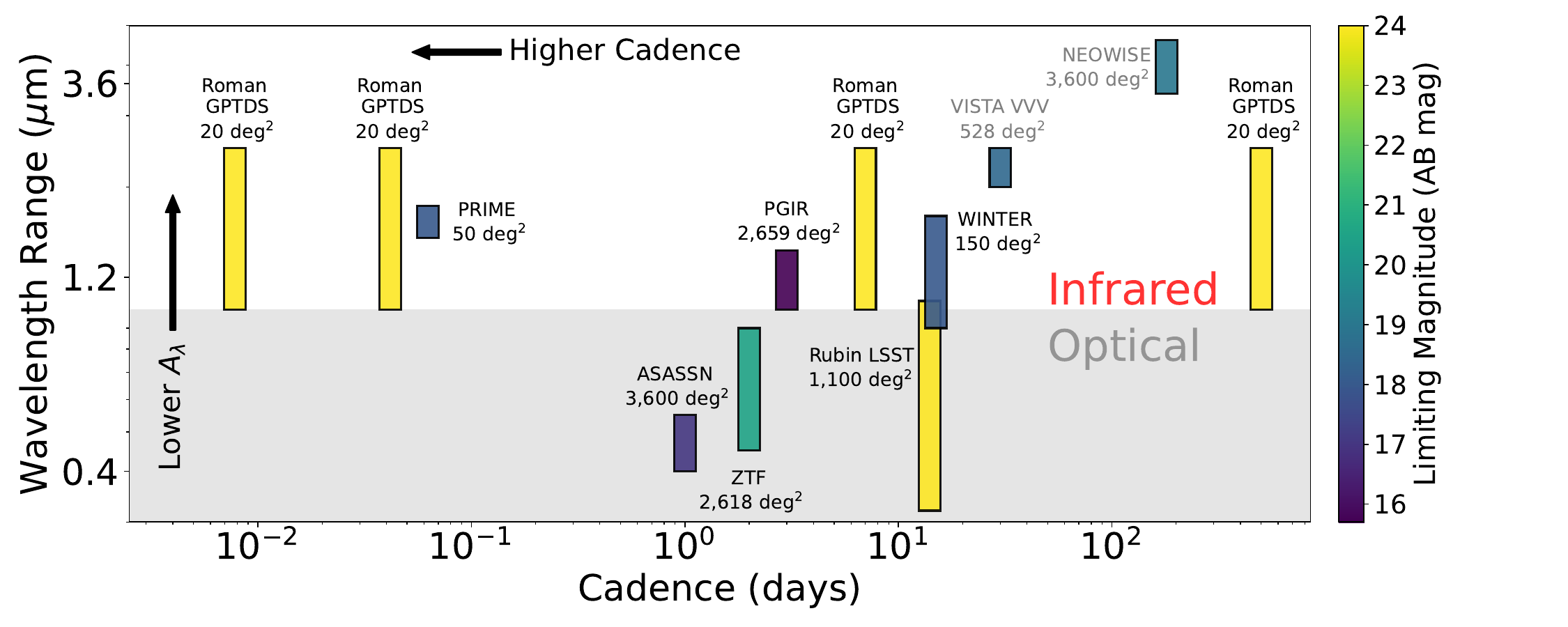}
    \caption{A comparison of WINTER to other past, present, and future optical and infrared surveys of the Galactic plane, defined for this figure as surveys between galactic latitude $|b| < 5^{\circ}$. Concluded surveys are shown in gray. In contrast to PGIR, PRIME, and VVV, WINTER has color information from three filters, allowing for studies of the physics underlying the outbursts from the evolution of the behavior across color bands. Surveys shown include the Roman Galactic Plane Time Domain Survey (GPTDS) \citep{roman_surveys}, PRIME \citep{PRIME_camera}, ASASSN \citep{ASASSN}, ZTF \citep{ZTF}, PGIR \cite{De:2020}, Rubin LSST Wide-Fast-Deep survey \citep{PSTN-056}, WINTER, VISTA VVV \citep{Minniti:2010}, and NEOWISE \citep{NEOWISE}.}
    \label{fig:winter_comp}
\end{figure*}

WINTER identified two new FUor outbursts that are bright in the near-infrared, also detected in the mid-infrared, but largely obscured at optical wavelengths. \1 is a previously characterized FUor and \2 is a newly classified FUor from this sample. Long-term monitoring with surveys like WISE and Gaia show the rise and color evolution of the bursts, revealing years-long events with slow declines. WINTER observations of the events confirmed that these objects were bright in the NIR and prompted rapid spectroscopic follow-up. The spectra for both events show cool objects with mixed spectral types with evidence of winds or outflows. Furthermore, the spectra confirm the events as FUors in the \citealt{Connelley:2018} classification scheme. Compared to the larger sample of FUors, the two WINTER events show notably high CO equivalent widths (Figure~\ref{fig:classification_yso}) and steep spectral slopes. 

Although \1 was previously reported, we present its Gaia, WINTER, and NEOWISE light curve and progenitor properties. This indicates that the mid-infrared rise seen in NEOWISE started shortly before ($\sim$1 year) the Gaia optical outburst. This is very similar to the object in \citealt{Guo:2024b}, which also shows a later time re-brightening as seen in this object. 

This paper presents the spectroscopic confirmation of \2 as a bona fide FUor, previously identified as a candidate based on its mid-infrared variability. Despite the lack of any optical signature during the outburst, NEOWISE photometry reveals that the event began around 2014 and has continued to brighten steadily in the infrared. Given the estimated distance and the high extinction inferred from SED fitting—well in excess of the expected foreground extinction—the source is likely a deeply embedded YSO. This level of obscuration naturally explains the absence of any optical counterpart throughout the duration of the outburst.

\begin{deluxetable}{lcc}
\tablecaption{Derived Parameters for WINTER FUor Outbursts.\label{tab:wntr_params}}
\tablehead{
\colhead{Parameter} & \colhead{WNTR24-cua} & \colhead{WNTR24-egv}
}
\startdata
Distance (kpc)      & 2.40         & 2.52 \\
$A_V$ (mag)         & 5.0         & 19.0 \\
$T_{\mathrm{eff}}$ (K) & 3000        & 4000 \\
$L_{\mathrm{pre}}$ ($L_\odot$) & 0.6        & 5.6 \\
$L_{\mathrm{post}}$ ($L_\odot$) & 5.5          & 222.1 \\
$L_\star$ ($L_\odot$)  & 0.033        & 2.97 \\
$M_\star$ ($M_\odot$)  & 0.15       & 0.60 \\
$R_\star$ ($R_\odot$)  & 0.86        & 2.00 \\
$\dot{M}$ ($M_\odot$ yr$^{-1}$) & $1.3\times10^{-6}$ & $3.0\times10^{-5}$ \\
\enddata
\end{deluxetable}

Though both objects exhibit spectroscopic similarity, their outburst timescales differ markedly. \1 shows a rapid rise, with the infrared brightening (seen in WISE $W2$ and VVV $Ks$) about a year before the optical Gaia $G$ outburst (Figure \ref{fig:WNTR24_lc}). In contrast, \2 is undergoing a slower outburst, with longer wavelengths likely leading shorter ones, as evidenced by W2 rising more gradually and persisting longer than W1. The color evolution of both sources aligns with outside-in propagation models, which predict the infrared to brighten before the optical. \citet{Cleaver:2023} show that the delay between infrared and optical rise correlates with accretion rate, with more luminous (i.e., higher accretion rate) events exhibiting longer delays. It is therefore possible that \2 has not yet reached its optical peak, and the current infrared brightening is an early precursor—as some models predict the IR outburst can precede the optical by decades \citep[e.g.,][]{Cleaver:2023}.

Overall, these studies highlight the importance of multi-color monitoring of the Galactic plane to understand the triggering mechanisms of FUors. The classical thermal instability model predicts that the IR rise lags behind the optical \citep{Bell:1995}. In contrast, magnetorotational instability (MRI) models predict an infrared precursor, with the delay depending on the luminosity and mass accretion rate. For the largest outbursts---such as FU Orionis itself---\citet{Cleaver:2023} argue that the outburst originates far out in the disk, producing an infrared precursor decades before the optical rise. However, the delay is shorter for smaller peak outbursts, where the instability occurs closer to the star. Models by \citet{Nayakshin:2024} instead suggest that the evaporation of a close-in planet can trigger FUor outbursts; in this case, the delay does not depend on the mass accretion rate. \citet{Guo:2024b} find that such a scenario is more consistent with similar delays observed in outbursts with varying peak accretion rates. 

While models disagree on whether the IR–optical delay should scale with accretion rate, distinguishing between these scenarios requires monitoring across multiple wavelengths and cadences from weeks to decades. Prior infrared surveys like VVV and PGIR lacked the multi-color, high-cadence coverage needed to resolve early outburst evolution, while NEOWISE operated only at mid-infrared wavelengths where spectral energy distributions are less sensitive to changes in temperature. Optical time-domain surveys are largely blind to these events due to extinction. WINTER’s near-infrared capabilities—combining $Y$, $J$, and $H_s$ band coverage with regular cadence—directly address this gap, offering sensitivity to the color evolution that discriminates between theoretical models (e.g., inside-out versus outside-in instability propagation; \citealt{Bell:1995, Cleaver:2023}). Although WINTER observations of the two FUors studied here began only after their outbursts were underway, future real-time detections will enable systematic tests of these models.

When the Galactic plane is visible, WINTER continues its survey of the plane and nearby star-forming regions, enabling ongoing monitoring of the two FUor events presented here, as well as the discovery of new outbursts. The survey operates on a two-week cadence in $J$- and $Hs$-bands, targeting northern fields including active star-forming regions such as W40 and the North America Nebula. Near-IR coverage will catch obscured events missed by optical surveys and help fill in an infrared gap now that mid-IR monitoring with WISE has concluded. In particular, WINTER is prioritizing coverage of the northern fields selected for the upcoming Roman Galactic Plane Survey, providing long-term near-infrared monitoring ahead of Roman’s deeper surveys. Roman observations will also provide resolved follow-up of outbursts seen in WINTER. 

The 2-week cadence and time-domain infrastructure of WINTER enable rapid spectroscopic follow-up of newly detected outbursts, allowing classification of short-duration bursts or while long-duration events are still evolving (Figure \ref{fig:winter_comp}). Early spectroscopy is particularly valuable for capturing transient features during the rise phase, when the structure of the accretion disk is still forming. Timely classification also helps distinguish between different types of transients—such as FUors, EXors, classical novae, or planetary engulfment events \citep{Audard:2014, Chomiuk:2021, De:2023}. In addition, the high cadence improves sensitivity to short-duration, low-amplitude events that are easily missed by slower surveys. WINTER is uniquely positioned to detect and spectroscopically characterize both short-lived transients and long-duration FUor-type events.

As both NEOWISE and VVV have now concluded operations, WINTER will provide ongoing infrared coverage of the Galactic plane in the years leading up to the launch of the Roman Space Telescope and its planned Galactic plane survey (RGPS) \citep{roman_surveys}. WINTER’s high-cadence, multi-band near-infrared monitoring complements the deep optical observations from the Vera C. Rubin Observatory \citep{Rubin}, the high-resolution infrared imaging expected from Roman, and deep single-band NIR monitoring from the future Cryoscope survey \citep{Kasliwal:2025}. Together, these facilities will enable more complete identification and characterization of YSO outbursts, spanning a wide range of evolutionary stages and environments. In particular, WINTER’s early detections will ensure that FUors erupting during or after the RGPS can be linked to Roman photometry of their faint progenitors---for example, the progenitor of \1 was barely detected in pre-outburst imaging. Additionally, for active FUors, Roman will also offer the opportunity to study spatially resolved outflows, complementing what has previously only been achievable through targeted Hubble Space Telescope follow-up. As WINTER continues its Galactic plane survey, it is building a growing sample of eruptive YSOs—including both classical FUors and more rapidly evolving events—that will serve as a valuable resource for multi-wavelength studies of episodic accretion in the time-domain era.

\section{Acknowledgments}
WINTER’s construction is made possible by the National Science Foundation under MRI grant number AST-1828470 with early operations supported by AST-1828470. Significant support for WINTER also comes from the California Institute of Technology, the Caltech Optical Observatories, the Bruno Rossi Fund of the MIT Kavli Institute for Astrophysics and Space Research, the David and Lucille Packard Foundation, and the MIT Department of Physics and School of Science. D.F.'s contribution to this material is based upon work supported by the National Science Foundation under Award No. AST-2401779. This research award is partially funded by a generous gift of Charles Simonyi to the NSF Division of Astronomical Sciences. The award is made in recognition of significant contributions to Rubin Observatory’s Legacy Survey of Space and Time. Furthermore, we acknowledge the support of the National Aeronautics and Space Administration through ADAP grant number 80NSSC24K0663. This work is based on observations obtained at the Infrared Telescope Facility, which is operated by the University of Hawai‘i under contract 80HQTR19D0030 with the National Aeronautics and Space Administration. We acknowledge ESA Gaia, DPAC and the Photometric Science Alerts Team (http://gsaweb.ast.cam.ac.uk/alerts). This research made use of hips2fits,\footnote{https://alasky.cds.unistra.fr/hips-image-services/hips2fits} a service provided by CDS. Any use of generative AI in this manuscript adheres to ethical guidelines for use and acknowledgment of generative AI in academic research. Each author has made a substantial contribution to the work, which has been thoroughly vetted for accuracy, and assumes responsibility for the integrity of their contributions.

\bibliography{winter}{}
\bibliographystyle{aasjournal}

\end{document}